\documentstyle[12pt,epsf]{article} 
\newlength{\abstractwidth} 
\renewcommand{\thefootnote}{\fnsymbol{footnote}} 
\renewcommand{\thanks}[1]{\footnote{#1}} 
\newcommand{\starttext}{ 
\setcounter{footnote}{0} 
\renewcommand{\thefootnote}{\arabic{footnote}}} 
\renewcommand{\theequation}{\thesection.\arabic{equation}} 
\newcommand{\be}{\begin{equation}} 
\newcommand{\bea}{\begin{eqnarray}} 
\newcommand{\eea}{\end{eqnarray}} 
\newcommand{\beq}{\begin{equation}} 
\newcommand{\ee}{\end{equation}} 
\newcommand{\eeq}{\end{equation}}

\def\ba{\begin{eqnarray}} 
\def\ea{\end{eqnarray}}

\def\12{{1 \over 2}} 
\def\32{{3 \over 2}} 
\def\72{{7 \over 2}} 
\def\92{{9 \over 2}}

\def\nc{non--commutative} 
\def\ny{non--commutativity} 
\def\snc{space/time non--commutative} 
\def\sny{space/time non--commutativity} 

\begin{document} 
\renewcommand{\theequation}{\thesection.\arabic{equation}} 
\begin{titlepage} 
\bigskip 
\rightline{IASSNS-HEP-00/34} 
\rightline{SU-ITP 00-14} 
\rightline{hep-th/0005015}

\bigskip\bigskip\bigskip\bigskip

\centerline{\Large \bf {Space/Time Non--Commutativity and Causality }}

\bigskip\bigskip 
\bigskip\bigskip 
\centerline{\it N. Seiberg,} 
\medskip 
\centerline{School of Natural Sciences} 
\centerline{Institute for Advanced Study} 
\centerline{Olden Lane, Princeton, NJ 08540} 
\bigskip

\centerline{\it L. Susskind and N. Toumbas} 
\medskip 
\centerline{Department of Physics} 
\centerline{Stanford University} 
\centerline{Stanford, CA 94305-4060} 
\bigskip\bigskip 
\begin{abstract} 
\noindent 
Field theories based on \nc \ spacetimes exhibit very distinctive 
nonlocal effects which mix the ultraviolet with the infrared in 
bizarre ways. In particular if the time 
coordinate is involved in 
the \ny \ the theory seems to be seriously acausal and inconsistent 
with conventional Hamiltonian evolution. To illustrate these 
effects we study the scattering of wave packets in a field theory 
with space/time \ny. In this theory we find effects which seem to 
precede their causes and rigid rods which grow instead of Lorentz 
contract as they are boosted. These field theories are evidently 
inconsistent and violate causality and unitarity.

\noindent 
On the other hand open string theory in a background electric field 
is expected to exhibit space/time \ny. This raises the question of 
whether they also lead to acausal behavior. We show that this is 
not the case. Stringy effects conspire to cancel the acausal 
effects that are present for the \nc \ field theory.

\medskip 
\noindent 
\end{abstract} 
\end{titlepage} 
\starttext \baselineskip=18pt \setcounter{footnote}{0}

\setcounter{equation}{0} 
\section{Introduction}

Non-commutative field theory 
is a model of a world with non-commuting spatial coordinates 
(space/space non--commutativity) and in fact such field theories 
do arise as the description of string theory in certain 
backgrounds \cite{1,2,3}. Non--commutative theories are very nonlocal in the 
non-commuting spatial directions but are quadratic in time 
derivatives. Nonlocality in spatial directions ruins Lorentz 
invariance but it is consistent with the basic rules of 
Hamiltonian quantum mechanics. Action at a 
distance may occur but events never precede their causes.

The situation is much less clear for field theories with 
non--commutativity between time and a space direction (space/time 
non--commutativity). The action is arbitrarily non-local in time with 
the evolution of fields at one time depending on the value of fields 
at both past and future times. The question then is whether the kind of 
unusual behavior found in space/time \nc \ field theories 
can ever occur in any consistent theory with 
a Hamiltonian and a unitary S-Matrix. We don't know the answer to 
this question but we examine an obvious candidate, 
string theory in a background electric $B_{\mu \nu}$ 
field. This theory is manifestly unitary 
and may be expected to exhibit effects similar to those seen in the 
field theory example. 
However as we have seen in \cite{4,5} the theory in an electric 
field never becomes a field theory and retains its stringy 
excitations. We will see that the stringy effects cancel the 
acausal effects of space/time \ny .

The plan of the paper is as follows. In section 2 we 
describe the scattering of wave packets in \nc \ field theory. In 
the case of space/space \ny \ the scattering induces a sudden 
spatial displacement of the wave packet in the direction orthogonal 
to the momentum. The magnitude of displacement is proportional to 
the momentum. As expected the degree of nonlocality increases with 
momentum. The scattered particles behave like rigid rods oriented 
perpendicular to their momentum with a size proportional to their 
momentum \cite{3,6,7,8}.

In the space/time case the wave packets scatter in a manner that 
appears to violate causality. Two scattered packets appear. One of 
them is physically sensible and corresponds to a time delay 
proportional to the incoming momentum. The other wave packet 
has a negative time delay! 
As in the space/space case 
the effect increases with the momentum. Thus at very high energy 
one of the outgoing waves appears to originate long before the particles 
could have collided. We refer to this behavior as {\it{advanced}}. 
Alternatively the particles behave like rods 
oriented along the direction of motion. Again the length of the 
rod is proportional to its momentum. This increase of length with 
momentum is very counterintuitive and is quite opposite to the 
expected Lorentz contraction.

Having defined the distinctive signatures of space/time non--commutativity, 
in Section 3 we proceed to look for string theory realizations of 
these signatures. We investigate open string scattering 
with and without background electric fields. We 
find delayed wave packets with time delay 
proportional to momentum as expected in \nc \ theories. We emphasize 
that the delayed effect occurs {\it{with and without}} 
a background electric field. 
However in no case do we find 
the acausal signatures of space-time \ny . The case of open 
strings in electric fields is particularly interesting. Although 
the amplitudes acquire Moyal phases the stringy effects mask the 
phases that would otherwise give rise to advanced effects. The 
scattering with the electric field does not seem appreciably 
different than without it even in the critical limit.


\setcounter{equation}{0} 
\section{Scattering in Non--Commutative Field Theory} 
In this section we study the effect of space/space and space/time 
non--commutativity on the scattering of massless scalar particles. 
We begin with the space/space case. To illustrate the main points 
it is sufficient to consider 2+1 dimensional \nc \ scalar $\phi^4$ 
theory in lowest order perturbation theory. The coordinates are 
labeled $(x,y,t)$. Since this case is familiar we will just 
describe the scattering schematically. Let us consider two high 
energy particles moving along the $x$ axis with spatial momentum 
$P_x$. We will take the initial wave function to be 
\be 
\Psi(x,y) = \exp(i P_x x) \psi_{in}(y) \ee where 
\be 
\psi_{in}(y) = \int dP_y \hat{\psi}_{in}(P_y) e^{i P_y y}. \ee The 
important feature of the scattering amplitude for our purposes is 
the Moyal phase factors which take the form 
\be 
M= \exp {{i\over 2} \theta(P_x Q_y -P_y Q_x)} \ee where $Q$ is the 
momentum transfer. We assume $P_x \gg P_y, Q_x, Q_y$.

After the scattering, the scattered momentum space wave function 
is given by an expression of the form 
\be 
\hat{\psi}_{out}(P_y)=\int dQ_y \hat{\psi}_{in}(P_y+Q_y) \exp 
{{i\over 2}\theta (P_x Q_y)}. \ee

In coordinate space 
\be 
\psi_{out}(y) \approx \psi_{in}(y) \delta(y -{1 \over 2}\theta 
P_x). \ee In other words the outgoing scattered wave appears to 
originate from the displaced position $y =\theta P_x/2 $.

An intuitive way to understand this effect is to think of the 
incident particles as extended rods oriented perpendicular to 
their momentum \cite{3,6,7,8}. The size of the rods is $\theta P$ and 
the rule is that they only interact if their ends touch.

Now we turn to the more interesting case of space/time \ny \ which 
we will study in much more detail. For simplicity we will work in 
$1+1$ dimensions. We denote time by $t$ and the spatial variable 
by $x$.

Let us begin by reviewing the scattering of wave packets in 1+1 
dimensions. A free scalar field in $1+1$ dimensions has the 
following Fourier decomposition 
\be 
\phi(x, 0) = \int {dp \over (2\pi) \sqrt{2E_p}}\left(a_p e^{ipx} + 
{a}_p^{\dagger} e^{-ipx}\right), \ee with 
\be 
\left[a_p, a_k^{\dagger}\right]= (2\pi)\delta(p - k). \ee 

Because of the special infrared divergences of massless 1+1 
dimensional scalar fields we will work with the derivative of 
$\phi$ rather than $\phi$ itself: 
\be 
\phi'(x, 0) =i \int {dp \over (2\pi) \sqrt{2E_p}}\left(p a_p 
e^{ipx} 
- 
p {a}_p^{\dagger} e^{-ipx}\right). \ee

Single particle states with momentum $p$ are normalized as follows 
\be 
|p> = \sqrt{2E_p} a_p^{\dagger} |0>. \ee Then the norm 
\be 
<p|k> = 2 E_p (2\pi) \delta(p - k) \ee is Lorentz invariant. The 
wavefunction of such a state will be defined by 
\be 
<0|\phi'(x)|p> =ip e^{ipx}. \ee Using the equation of motion for 
the free scalar field, we can find the wavefunction at all times.

Next, we turn on some interactions. For example, consider a commutative 
$\phi^4$ interaction. We are interested in the scattering of 
massless scalars, in particular 2--body to 2--body scattering. For 
sufficiently high energies, we can use perturbation theory to 
calculate an S--matrix. The S--matrix takes the following form 
\be 
S = 1 + iT, \ee where 
\be 
<p_1, p_2|iT|k_1, k_2> = (2\pi)^2 \delta^2(k_1 + k_2 - p_1 - 
p_2)i{\cal{M}}(k_1, k_2 \rightarrow p_1, p_2). \ee Here, $k_1, 
k_2$ denote the 2--momenta of the incoming particles and $p_1, 
p_2$ the 2--momenta of the outgoing particles. The invariant 
amplitude $i{\cal{M}}$ is computed in the usual way using Feynman 
diagrams. For the simple case of a $\phi^4$ interaction, 
\be 
i{\cal{M}} = -ig \ee to leading order in perturbation theory. In 
$1+1$ dimensions the only effect one expects to see in 2--body to 
2--body scattering is time delays.

Now, consider an incoming state consisting of correlated pairs of 
particles with opposite momenta: 
\be 
|\phi>_{in}= \int {dk \over (2\pi){2E_k}}\phi_{in}(k)|k, -k>, \ee 
with 
\be 
\phi_{in}(k) = \phi_{in}(-k). \ee The wavefunction of such a state 
is given by 
\be 
\Phi_{in}(x)\equiv <0|\phi'(x_1)\phi'(x_2)|\phi>_{in} = 2 \int { dk k^2 \over 
(2\pi){2E_k}}\phi_{in}(k)e^{ikx}, \ee where $x = x_1 - x_2$ is the 
relative separation of the two particles. There is no dependence 
on the center of mass position, since the overall center of mass 
momentum is zero. Let us also choose $\phi_{in}(k)$ so that at the 
time of the collision $t=0$, the wave-packet is well concentrated 
at $x = 0$. Then the incoming particles are close together at 
$t=0$. For example, we may choose 
\be 
{\phi_{in}(k)} = e^{-{(k - k_0)^2 \over \lambda}} + e^{-{(k + k_0)^2 
\over \lambda}}. \ee The wavepacket is concentrated at energies closed 
to $k_0$. The width of the packet in space is given by $1/\lambda^{1/2}$. 
We let 
$\lambda \ll k_0^2$ and take $k_0$ large. At earlier times, 
$t<0$, we can use the free equations of motion to find that the 
packet is concentrated at $x = 2t$. This means that the incoming 
particles are far apart in the past and they collide at $t=0$.

Similarly, the outgoing state is taken to be 
\be 
|\phi>_{out}= \int { dp \over (2\pi){2E_p}}\phi_{out}(p)|p, -p>. 
\ee Then, 
\be 
|\phi>_{out}= S|\phi>_{in}=|\phi>_{in} + iT|\phi>_{in}. \ee 
Therefore, we have that 
\be 
{\phi_{out}(p) \over (2\pi){2E_p}} = {\phi_{in}(p) \over 
(2\pi){2E_p}} + {<p, -p|iT|\phi>_{in} \over 8(2\pi)^2 E_p^2 
\delta(0)}. \ee Now, using the form of the matrix element $<p, 
-p|iT|k, -k>$, eq. (2.13), we find that the non--trivial part of 
$\phi_{out}(p)$ is given by 
\be 
\int{dk \over (2\pi){2E_k}}\phi_{in}(k)({i{\cal{M}} \over 
8E_p^2})\delta(2E_k - 2E_p)={\phi_{in}(p) \over (2\pi){2E_p}} 
{i{\cal{M}} \over 8E_p^2}. \ee Therefore, using eq. (2.17), the 
non--trivial part of the outgoing wavefunction can be obtained by 
\be 
\Phi_{out}(x)\equiv <0|\phi'(x_1)\phi'(x_2)|\phi>_{out}= 2 \int {dp p^2 \over 
(2\pi){2E_p}}\phi_{in}(p){i{\cal{M}} \over 8E_p^2}e^{ipx}. \ee

In the case of the $\phi^4$ theory, we see that nothing much 
happens. Choose $\phi_{in}(p)$ to be a polynomial in $p$ times a 
Gaussian so that the integral converges. Then $\Phi_{in}(x)$ is 
concentrated at $x=0$. Since $i{\cal{M}} \sim g$, at time $t=0$, the outgoing 
wavefunction will also be concentrated at $x = 0$. 
Therefore, there are no large time delays. Using the free 
equations of motion, we find that at later times the wave-packet 
is concentrated at $x = 2t$ and so the outgoing particles separate 
in the far future.

Consider now the effect of space/time \ny . 
\be 
[t, x]=i\theta. 
\ee 
The theory is defined by replacing the ordinary product by 
a $*$--product given by 
\be 
\phi_1*\phi_2(x, t)=e^{i{\theta \over 
2}[\partial_0^{y}\partial_1^{z}-\partial_1^{y}\partial_0^ 
{z}]}\phi(y)\phi(z)|_{y=z=(x, t)}. \ee The $\phi^4$ Lagrangian 
contains now an infinite number of time derivatives from the 
interaction term 
\be 
g\phi*\phi*\phi*\phi. \ee Therefore the theory is not local in 
time. It is not clear that such a theory has a well defined 
Hamiltonian. One plus one dimensional Lorentz invariance however, 
is undisturbed by the \ny. This is easily seen from the fact that 
the defining commutation relation has the form 
\be 
[x^{\mu }, x^{\nu}]=i\theta \epsilon^{\mu \nu}. \ee

The effect of the $*$--product is to produce phases in the 
interaction vertex that depend on the energies of the particles. 
The tree--level scattering amplitude is now given by 
\be 
i{\cal{M}} \sim g[\cos(p_1\wedge p_2)\cos(p_3 \wedge p_4) + 
2\leftrightarrow 3 + 2\leftrightarrow 4], \ee where $p_1, p_2, p_3, p_4$ 
are the 2--momenta of the particles satisfying 
\be 
p_1 + p_2 + p_3 + p_4 = 0. \ee Here $p\wedge k=\theta 
(p^0k^1-k^0p^1)$. 
Note that we have used conventions with all particles 
taken to be incoming in the vertices; i.e. energies of outgoing 
particles are negative. 
In the center of mass frame with the incoming 
particles (and outgoing) having equal and opposite spatial momenta, the 
amplitude becomes 
\be 
i{\cal{M}} \sim g[\cos(4p^2\theta)+2]. \ee The pattern is similar 
in more general \nc \ theories but depending on the spins and 
polarizations of the particles 
the periodic functions may be sines in place of cosines.

We remark that such a theory fails to be unitary at the 1--loop 
level \cite{9}. However, let us just consider tree--level scattering 
amplitudes and in particular the effect of non--commutativity on 
the outgoing wave-packets. We choose for the incoming wave-packet 
$\phi_{in}(p)$ a gaussian function: 
\be 
{\phi_{in}(p)} \sim E_{p}\left( e^{-{(p - p_0)^2 \over \lambda}} 
+e^{-{(p + p_0)^2 \over \lambda}}\right). \ee (The extra factor of 
$E_p$ is added to simplify the integrals but does not change the 
qualitative behavior of our results.) Using 
eq. (2.19), we can find the outgoing wavefunction 
\be 
\Phi_{out}(x) \sim g \int dp[\cos(4p^2\theta)+2]\left(e^{-{(p - 
p_0)^2 \over \lambda}} + e^{-{(p + p_0)^2 \over \lambda}}\right)e^{ipx}. 
\ee To 
compute the integral, we need to calculate the following Fourier 
transform

\bea 
& &\int dpe^{4ip^2\theta}e^{-{(p - p_0)^2 \over \lambda}}e^{ipx} \sim 
e^{{-p_0^2 \over \lambda}}{1 \over \sqrt{{1 \over \lambda} - 
4i\theta}} e^{{({2p_0 \over \lambda} + ix)^2 {1\over 4({1 \over 
\lambda} - 4i\theta)}}} \cr &=& 
{1 \over \sqrt{{1\over \lambda}-4i\theta}} \exp {\left[- \lambda 
\left( x+ 8p_0 \theta \right)^2 \over 4(1+16 \theta^2 \lambda^2) 
\right]} \exp {\left[ -i \theta \lambda^2 \left( x- {p_0 \over 
2\lambda^2 \theta} \right)^2 \over 1+16 \theta^2 \lambda^2 
\right]} \exp{i{p_0^2 \over 4 \lambda^2 \theta}}. \eea 
We take $p_0 \gg 
\lambda^{1/2} \gg 1/p_0\theta$, and also assume that $\lambda \theta 
\gg 1$. Then, 
eq. (2.33) simplifies as follows 
\be 
{1 \over \sqrt{ - 4i\theta}} e^{- {(x + 8p_0\theta)^2 \over 
64\theta^2\lambda}}e^{-i {(x - {p_0 \over 2\lambda^2 \theta})^2 \over 
16\theta}}e^{i{p_0^2 \over 4 \lambda^2 \theta}} \equiv 
F(x;\theta,\lambda,p_0). \ee Then the outgoing wavefunction is 
given by 
\be 
\Phi_{out}(x) \sim g \left[ F(x;-\theta,\lambda,p_0)+ 
4 \sqrt{\lambda}e^{-\lambda {x^2 \over 4}}e^{ip_0 x} + 
F(x;\theta,\lambda,p_0) \right] + 
(p_0 \rightarrow -p_0) . \ee

We see that the wave-packet splits into three parts, one 
concentrated at $x = 8p_0\theta$, one at $x=0$ and the other at 
$x = -8p_0\theta$. The width of the first and third packet is given by 
$8\lambda^{1/2}\theta$ while the one concentrated at $x=0$ has 
width $2/\lambda^{1/2}$. Therefore, the packets are well separated 
for $p_0 \gg \lambda^{1/2} \gg 1/p_0\theta$. The separation of the two 
displaced 
packets is proportional to $p_0$ which is the energy of the 
particles. The bigger the energy is the bigger the separation.

The packet at $x=0$ oscillates with frequency $p_0$. The other two 
packets oscillate with phases $\exp[i(x + p_0/2\lambda^2 
\theta)^2/16\theta]$ and $\exp[-i(x - p_0/2\lambda^2 
\theta)^2/16\theta]$. 
Locally, near the 
maxima at $x=\pm8p_0\theta$ the phases in the other two packets become 
$\exp[ip_0(1+1/16\lambda^2\theta^2)\Delta x]$ and so they oscillate 
with frequency $p_0$ since $\lambda \theta \gg 1$. 
This was expected from energy conservation.

All three wavepackets propagate towards $x \rightarrow 
\infty$. They correspond to particles $3$ and $4$ moving apart. In 
our conventions, particle $4$ has momentum opposite to that of 
particle $1$. We can think of it as particle $1$ back--scattered. The 
first packet is an advanced wave. It appears at $x=0$ at some time 
before the incoming wave arrives at the origin. The phase responsible 
for the acausal behavior is $e^{-4i\theta p^2}$. The third packet is 
delayed. The opposite phase causes the delay. Similarly the terms we 
get from $p_0 \rightarrow -p_0$ are waves moving towards $x 
\rightarrow -\infty.$ It is easy to check that the advanced wave is 
again produced by the phase $e^{-4i\theta p^2}$.

Thus the collision is described as follows: The center of mass back 
scattering is isomorphic to bouncing off a wall. An incoming wave 
packet of spatial width $\lambda^{-1/2}$ is arranged to arrive at 
the wall at time $t=0$. The outgoing wave consists of three terms. 
One term appears to originate from the wall at time $t=8p_0 
\theta$, well after the incoming packet reached the wall. It is 
odd that the wave is delayed for so long a time as the energy 
increases but it is not acausal. A second term is neither 
significantly delayed or advanced. We will ignore it.

The other term is an ``advanced'' wave which appears to leave the 
wall {\it{before}} the incoming packet arrived. What is worse, the 
effect increases with energy so that the advance is proportional 
to the energy. This certainly seems acausal.

In itself, an advance does not violate causality. 
A simple non-relativistic model 
illustrates the point. Picture the incoming particles as rigid rods of 
length $L$. Assume the rod reflects when its leading end strikes 
the wall. In this case the center of mass of the rod will appear 
to reflect before it reaches the wall. In a Newtonian world a 
physicist measuring an advance would conclude that the scattering 
objects resembled rigid rods.

The problem with such rigid rods is that they conflict with 
the combined constraints of causality and Lorentz invariance. In 
fact the required properties of the rod are completely at variance 
with the usual expectations of special relativity. For example one 
usually assumes that perfectly 
rigid bodies can not exit. The reason is that by suddenly 
displacing one end of a rod, the 
signal would instantly appear at the other end. Since such a rod is 
spacelike, this is usually thought to lead to action at a distance, 
nonlocality and violation of causality.

Equally peculiar is the behavior of the rods under boost. 
Suppose the 
momentum is increased. The conventional expectation is that the 
rod will Lorentz contract thus decreasing the advance. This is 
precisely the opposite of what space/time \ny \ implies. The rod 
seems to expand as its momentum increases.

Another phenomenon predicted by eq. (2.35) is that the outgoing packet 
is much broader than the incoming. Let the incoming packet be of 
spatial width $\lambda^{-{1\over 2}}$. By contrast, the outgoing 
packet has spatial width $\lambda^{1\over 2}\theta$. In the limit we 
study of large $\lambda \theta$ this is broader than the incoming 
packet. How is this explained?



To understand this effect we return to the rod model. The advance 
is of order the rod size $L$. If we take $L = p\theta$ then the 
uncertainty in the rod size is 
\be 
\Delta L = \theta \Delta p = \theta \lambda^{\12}. \ee This means 
that the advance is also uncertain by the same amount. This 
obviously broadens the outgoing packet by the required amount.

All three terms in eq. (2.35) can be interpreted in terms of the rod 
model. Each of the incoming rods has two ends, a leading and a 
trailing end. The advanced term is due to the scattering of the 
two leading ends while the retarded contribution originates from 
the interaction of the trailing ends. The interaction of a leading 
and a trailing end contributes the second term in eq. (2.35).

What are we to make out of this behavior? The most obvious 
response is to dismiss it as pathological and declare space/time 
\ny \ to be unphysical. Our opinion is that this is prematurely 
pessimistic. The main reason is that some of the properties of the 
amplitude largely follow from the uncertainty principle implied by 
eq. (2.24) 
\be 
\Delta t \Delta x \geq \theta. \ee This uncertainty principle has 
the same form as the stringy uncertainty principle \cite{12,13,14} 
\be 
\Delta t \Delta x \geq \alpha'. \ee It therefore behooves us to 
inquire into the structure of string theory amplitudes to see if 
they produce any behavior similar to what we find in theories 
with space/time \ny.


\setcounter{equation}{0} 
\section{Scattering in Open String Theory}

In this section, we analyze tree level scattering amplitudes of open 
strings on branes in the presence of a background electric $B_{\mu \nu}$
field. 
In the presence of a background electric field the underlying 
spacetime is non--commutative. It is interesting to ask how 
scattering experiments similar to those studied in the previous 
example can probe the space/time non--commutativity. In particular we 
would like to investigate whether the acausal behavior we found in the 
simple field theory model is present. As we shall see, the amplitudes 
produce causal behavior and exhibit large time delays proportional to 
the momentum. The later phenomenon persists even without the electric field. 
The reader may think of the problem in a $1+1$ dimensional context by 
considering open string scattering on a stack of D1-branes \cite{10,11}. 
Throughout this section, we denote by $g_s,l_s$ the string coupling 
constant and length scale.

At the level of disc amplitudes the inclusion of the electric field is 
simple. All we need is to start with the 
amplitudes at $E=0$, replace the metric $\eta_{\mu\nu}$ by the 
effective open string metric $G_{\mu\nu}$, $g_s$ by $G_s$ and multiply 
the answer by the phase factors with non--commutativity parameter 
$\theta$. In terms of the electric field these parameters are given by 
\cite{4}

\bea 
\cr G_{\mu\nu}=(1-\tilde{E}^2)\eta_{\mu\nu} \,,\ \mu\nu=0, 1 \cr 
G_{\mu\nu}=\delta_{\mu\nu} \,,\ \mu\nu \ne 0,1\cr 
\theta^{01} = 2\pi l_s^2 {\tilde{E} \over 1-\tilde{E}^2} \cr 
G_s = g_s (1-\tilde{E}^2)^{1 \over 2}. 
\eea 
Here, $\tilde{E} = E/E_{cr} \le 1$. The critical electric field is given 
by $E_{cr} = 1/2 \pi l_s^2$.

Let us consider the Veneziano amplitude describing massless open 
string scattering. In terms of open string parameters the amplitude has 
the following form: 
\bea 
{A_4} &\sim& G_s\left(K_{st}e^{i(p_1\wedge p_2 + 
p_3\wedge p_4)}+K'_{st}e^{ i(p_1\wedge p_4 + p_3 
\wedge p_2)}\right) 
{\Gamma(-2sl_s^2)\Gamma(-2tl_s^2) \over \Gamma(1+2ul_s^2)} \cr &+& 
G_s\left(K_{su}e^{i(p_1\wedge p_2 + 
p_4\wedge p_3)}+K'_{su}e^{i(p_1\wedge p_4 + 
p_2\wedge p_3)}\right) {\Gamma(-2sl_s^2)\Gamma(-2ul_s^2) \over 
\Gamma(1+2tl_s^2)} \cr &+& 
G_s\left(K_{tu}e^{i(p_1\wedge p_3 + p_2\wedge p_4)}+K'_{tu}e^{i(p_1\wedge
p_3 + p_4\wedge p_2)}\right) 
{\Gamma(-2tl_s^2)\Gamma(-2ul_s^2) \over \Gamma(1+2sl_s^2)}. 
\eea 
The amplitude $A_4$ is obtained by integrating four vertex 
operators around the disc. We denote the two incoming particles 
by 1 and 2 and the two outgoing particles by 3 and 4 and let all 
momenta be incoming. Using Mobius invariance the vertex operators 
of particles 1, 2 and 3 can be put at three fixed points on the 
boundary of the disc -- mapping it to the upper half plane these 
are usually taken to be $z_1=0$, $z_2=1$ and $z_3=\infty$ 
respectively. The location of the vertex operator of particle 
number 4, $z_4$, is then integrated over the real axis. Since a Mobius 
transformation does not change the cyclic ordering of the vertex 
operators, we need to add another piece obtained from fixing 
$z_4=1$ and integrating the location of particle number 2. The three 
terms in the answer correspond to $-\infty <z_4 <0$, 
$1<z_4<\infty$ and $0<z_4<1$ respectively and similarly for 
$2\leftrightarrow 4$. Here $p\wedge k = \theta^{01}(p_0k_1-k_0p_1)$.

The kinematic factors $K$ in (3.2) involve momenta, $p_i$, 
polarization vectors, $\xi_i$, and also traces over Chan Paton factors 
$\lambda_i$. 
The quantities $s,t,u$ are the Mandelstam variables 
\be 
s=2p_1p_2, \,\ t=2p_1p_4, \,\ u=2p_1p_3 \ee satisfying the mass 
shell constraint $s+t+u=0$. Scattering in the backward direction is 
defined by $u = 0$. In eq. (3.3) we used the open string metric to 
contract the indices.

For the case of backward scattering, $u=0$, the kinematics are such that 
only the first term corresponding to the $s$--channel exchange gets 
multiplied by phases. One phase occurs when particles $1$, $2$ and $3$ are 
placed at $z_1=0$, $z_2=1$ and $z_3=\infty$ respectively and the location of 
particle $4$ is integrated from $-\infty<z_4<0$. The opposite phase occurs
when 
$2 \leftrightarrow 4$. No phases multiply the other two terms. One of 
the two phases, $e^{-2\pi i \tilde{E}sl_s^2}$, caused the appearance 
of the advanced waves in the non--commutative field model.

Setting $u$ to zero, the first term in the amplitude takes the form 
\be 
A_{st} \sim G_s \left(K_{st}e^{2\pi i 
\tilde{E}sl_s^2}+K'_{st}e^{-2\pi i \tilde{E}sl_s^2}\right) 
\Gamma(-2sl_s^2)\Gamma(2sl_s^2). \ee 
Using the identity 
\be 
y\Gamma(y) \Gamma(-y) = -{\pi 
\over \sin(\pi y)}, \ee 
we can write this as follows 
\be 
A_{st} \sim G_s \left(K_{st}e^{2\pi i 
\tilde{E}sl_s^2}+K'_{st}e^{-2\pi i \tilde{E}sl_s^2}\right) 
{1 \over s \sin(2 \pi s l_s^2)}. \ee 
The 
kinematic factors $K$ are also simple in this case. They 
are proportional to $s^2$ times products of polarization 
vectors and traces over Chan Paton factors. Therefore, 
\be 
A_{st} \sim G_s s \left(a_1e^{2\pi i \tilde{E}sl_s^2}+ 
a_2e^{-2\pi i\tilde{E}sl_s^2}\right){1 \over 
\sin(2 \pi s l_s^2)}, 
\ee 
where the constants $a_1$ and $a_2$ are independent of $s$.

This term has poles at $s=n/2l_s^2$ with $n$ being an 
integer. The divergence of the amplitude at the poles is an essential 
physical feature of the amplitude, a resonance corresponding to the 
propagation of an intermediate string state over long spacetime 
distances. 
To define the poles we use the correct $\epsilon$ 
prescription replacing $s \rightarrow s + i\epsilon$. 
This has the effect of shifting the poles off the real axis. 
Then the function $1/\sin(2\pi s l_s^2)$ can be expanded as a power series in 
$y=e^{2i\pi sl_s^2-\epsilon}$. 
In all, this term in the amplitude takes the form 
\be 
A_{st} \sim G_s s \sum_{n >0 \,\ odd}a_1e^{2\pi i(n +\tilde{E}) 
sl_s^2} + a_2e^{2\pi i(n-\tilde{E}) s l_s^2} + O(\epsilon). 
\ee

Comparing with eq. (2.30), we see that the amplitude looks 
similar to the case of a non--commutative field theory. 
We get a sum of phases with the 
identification 
\be 
\theta'_n = 2\pi (n \pm \tilde{E}) l_s^2,\,\ n>0 \,,\ odd. \ee 
What is interesting is that the non--commutativity parameter $\theta$ 
gets modified by stringy oscillator effects. In fact the phases 
persist even in the absence of the electric field. We see that 
$\theta'_n$ are positive for all positive odd integers.

In contrast with the field theory case, here we get phases that cause 
time delays only. The ``acausal'' phase, $e^{-2\pi i\tilde{E}s}$, gets 
multiplied by powers of $y$ from the Gamma functions. 
The net effect is to produce phases which, for $\tilde{E}< 1$, or for 
$E< E_{cr}$, cause time delays. Evidently, even in the presence of a 
background electric field, string scattering amplitudes 
produce causal behavior only. The acausal behavior due to the 
non--commutativity parameter $\theta$ is cancelled by phases from the 
Gamma functions. It seems that the oscillators are crucial for the 
causal behavior of the theory. The effects of the non--commutativity 
are always mixed with the effects of the string oscillators. We see 
another reason why space/time non--commutative field theories cannot 
be obtained as limits of string theory in background electric fields, 
as was found in \cite{4,5}. 
Such theories show pathological acausal behavior and are not unitary. 
What is interesting, however, is that the onset of the acausal 
behavior we found occurs as the electric field approaches its critical value.

The other two terms in formula (3.2) can be analyzed in a similar 
way. It is easier to write 
\be 
A_{su}+A_{tu}=G_s(A_1+A_2), 
\ee 
where 
\bea 
A_1 \sim \left(K_{su}+K_{tu}\right) \left({\Gamma(-2sl_s^2)\Gamma(-2ul_s^2)
\over 
\Gamma(1+2tl_s^2)} + 
{\Gamma(-2tl_s^2)\Gamma(-2ul_s^2) \over \Gamma(1+2sl_s^2)}\right) 
\eea 
and 
\bea 
A_2 \sim \left(K_{su}-K_{tu}\right) \left({\Gamma(-2sl_s^2)\Gamma(-2ul_s^2)
\over 
\Gamma(1+2tl_s^2)} - 
{\Gamma(-2tl_s^2)\Gamma(-2ul_s^2) \over \Gamma(1+2sl_s^2)}\right). 
\eea 
Then we can analyze the sum and differences of the two combinations of Gamma 
functions that appear in (3.2) as $u \rightarrow 0$. Again, no Moyal 
phases multiply these two terms.

Setting $u=0$, we find that 
\be 
A_1 \sim a_3 s {\cos(2 \pi s l_s^2) \over \sin(2\pi s l_s^2)}. 
\ee 
Shifting $s \rightarrow s+i\epsilon$, we can expand $1/\sin(2\pi s 
l_s^2)$ in powers of $y$. Thus we find 
\be 
A_1 \sim a_3 s (1 + e^{4\pi i s l_s^2})\sum_{n\ge0\,\ even}e^{2\pi 
i n s l_s^2}. 
\ee 
The first term in the series is just proportional to $s$ and 
produces no large time delays. The other terms are phases responsible 
for time delays. The phases in this term are independent of $\theta$. 
They are present even in the absence of a background field. 
Ordinary scattering of open strings shares 
features with scattering in non--commutative field theory with 
effective non--commutativity parameter the string length 
squared. However the amplitude produces only causal behavior.

The other term produces a pole at $u=0$. We have that 
\be 
A_2 \sim a_4 s {1 \over ul_s^2 + i\epsilon}. 
\ee 
The pole in $u$ corresponds to the exchange of a massless particle and 
we will ignore it. This term has no oscillations in $s$.

The effect on the outgoing wave-packet is similar to the previous 
example except that the advanced waves are absent. 
Let us consider the case 
of no electric field for simplicity. Then $\theta'_n=2\pi n l_s^2$ 
and the open string metric $G_{\mu\nu}=\eta_{\mu\nu}.$ 
If we use 
eq. (2.23) for the same $\phi_{in}(p)$ given in eq. (2.31), we find 
for a typical phase in the amplitude 
\be 
\Phi_{out}(x) \sim G_s \int dp p^2 e^{4 i 
\theta'_n p^2}\left(e^{-{(p - p_0)^2 \over \lambda}} +e^{-{(p + p_0)^2 
\over \lambda}}\right)e^{ip x}. \ee For $p_0 \gg 
\lambda^{1/2} \gg 1/2\pi p_0 l_s^2$, this is proportional to

\be 
\Phi_{out}(x) \sim G_s {d^2 \over dx^2} 
F(x; 
\theta'_n,\lambda, p_0) + (p_0 \rightarrow -p_0) \ee with $ F(x; 
\theta_n,\lambda, p_0)$ given by eq. (2.34).

We see that the outgoing wave-packet splits into a series 
of packets, one localized at $x=0$, and a series at $x = 8p_0 \theta'_n 
\,\ n>0$. The advanced waves are absent. Only delayed waves are 
present. Each delayed packet has width given by 
$8\lambda^{1/2}\theta'_n$. 
The packets are not overlapping for $p_0 \gg \lambda^{1/2} \gg 
1/2\pi p_0 l_s^2$. The $n$--th packets are more spread. After the 
derivatives are performed we find that the contributions to the 
sum are dominated by the small $n$ packets. The amplitude of the packets 
falls like $n^{-5/2}$. Again the time delays are proportional to 
the energy $p_0$. It is interesting that the large time 
delays persist even in the absence of the electric field.

The interpretation is different than before. The scattering is 
causal. We would like to suggest the following to explain the series 
of time delays. As the two strings come together, an intermediate 
stretched string state is formed. The 
string state has total energy $p_0$. The state is oscillating from 
small size to a large size proportional to $p_0$. To see this we write 
\be 
p_0 = {L \over l_s^2} + {N \over L}, 
\ee 
where $N$ is some oscillation number. We see that this is minimized 
for $L \sim p_0l_s^2$. 
The state begins from 
small size and grows to a string of maximal size of order $p_0l_s^2$, 
storing the energy as potential energy. This repeats itself 
periodically. With each oscillation there is an amplitude for the 
string to split. Thus there is an infinite sequence of delayed 
wave packets. The delay is proportional to $L$ since the string ends 
move with the speed of light. The intermediate state has size 
proportional to the energy $p_0$. This is a manifestation of the 
stringy uncertainty relation.

In the case of a background electric field we find time delays (in 
closed string units) proportional to 
\be 
\Delta t = {p_0 l_s^2 \over 1 - \tilde{E}^2}. 
\ee 
The time delays are proportional to $1/T_{eff}$, where $T_{eff} = 
(1 - \tilde{E}^2) / l_s^2 $ 
is the effective tension of the open strings in the presence of the 
electric field. The effect of the electric field is to reduce the 
tension of the strings \cite{4,11}. As the 
electric field approaches its critical value, the time delays become 
longer. The extent of the intermediate state in space is also 
bigger. However, we note that as the field approaches its critical 
value, the effective coupling constant $G_s$ tends to zero and the 
amplitudes are suppressed.

We have illustrated the violations of causality in \snc \ 
field theory and its restoration in string theory by considering 
the evolution of wave packets. Evidently the scattering amplitudes 
of the field 
theory violate some principle of S-matrix theory that string 
theory preserves. In fact it is not difficult to see what 
principle is involved.
Macroscopic causality is usually assumed to follow from two 
properties of amplitudes. The first involves the location of 
singularities in the Mandelstam $s$ variable; namely, the 
amplitude should be analytic in the upper half plane. In the case 
of \nc \ field theory the amplitude in eq.(2.30) is an entire 
function and satisfies this rule. In the case of the string theory 
tree diagrams there is an infinite sequence of poles on the real 
axis. However with the conventional $i\epsilon$ prescription the 
poles are displaced to the lower half plane and lead to no 
violation of causality.
The second requirement is that the amplitudes should 
not exponentially 
diverge along any direction in the upper half plane in order 
to insure that certain contours of integration can be closed. 
This is what is violated in the \nc \ field theory. The cosine 
term in eq(2.30) exponentially diverges in the upper half plane. 
By contrast the \nc \ Moyal phases in string theory are 
compensated for by the factor $1/ \sin{(2\pi s l_s^2)}$ in 
formulae like eq(3.7) as long as the electric field is smaller 
than critical.

\setcounter{equation}{0} 
\section{Conclusion}

Space/time \ny \ is a more subtle phenomenon than its space/space 
counterpart. If we define the \snc \ deformation of a theory by 
multiplying its tree diagrams by space/time Moyal phases then an 
ordinary quantum field theory becomes acausal as well as 
non--unitary. The acausality is easily seen in the scattering of 
wave packets by the appearance of an outgoing signal that 
originates before the incoming particles reach each other.

By contrast, the \snc \ deformation of open string theory is not 
acausal. The theory does not have a limit in which stringy 
effects disappear. These stringy effects conspire to shift the 
Moyal phases so that they become causal. Thus the peculiar 
advanced effects found in the field theory should not be though of 
as the signature of \ny .

The delayed effects of \ny \ in a collision process 
are also interesting. The \sny \ 
manifests itself by time delays which grow linearly with 
increasing momentum. As we have seen in Section 3 the time delay 
of the leading delayed wave is governed by a parameter 
$\theta'_{\pm}=2 \pi (1\pm\tilde{E})l_s^2$. 
Nothing special seems to happen to $\theta'$ as the electric field 
is turned off. However $\theta'_{-} $ vanishes at the critical electric 
field. One possible interpretation of this is that open 
string theory exhibits the signature of 
\sny \ without any electric field. Indeed this interpretation is 
suggested by the well known space/time uncertainty principle 
\cite{12,13,14} i.e., it appears like the string grows in the 
longitudinal direction to a length of order $pl_s^2$.

\centerline{\bf Acknowledgements}

This paper is a revised version of an incorrect earlier version. 
We are very grateful to Igor Klebanov and Juan Maldacena for pointing 
out the error in the previous version. 
L.S. and N.T. would also like to acknowledge Igor Klebanov's help 
in analyzing the open string amplitudes in Section 3.

N.S. would like to thank the University of Texas Theory Group for hospitality 
during part of this work. We thank R. Gopakumar, S. Minwalla, 
D. Morrison, M. Peskin, S. Shenker and 
E. Witten for useful discussions. The work of N.S. was supported in 
part by DOE grant \#DE-FG02-90ER40542. The work of L.S. and N.T. was 
supported in part by NSF grant 980115.

\setcounter{equation}{0}

 

\begin{thebibliography}{99}

\bibitem{1} A. Connes, M.R. Douglas, A. Schwarz, 
``Noncommutative Geometry and Matrix Theory: Compactification on 
tori,'' hep-th/9711162, JHEP 9802:003,1998.




\bibitem{2} M.R. Douglas, C. Hull, 
``D--Branes And The Noncommutative Torus,'' 
hep-th/9711165, JHEP 9802:008,1998.




\bibitem{3} N. Seiberg, E. Witten, 
``String Theory and Noncommutative Geometry,'' hep-th/9908142, 
JHEP 9909:032,1999.




\bibitem{4} N. Seiberg, L. Susskind, N. Toumbas, ``Strings in 
Background Electric Field, Space/Time Noncommutativity and a new 
Non--Critical String Theory,'' hep-th/0005040.


\bibitem{5} R. Gopakumar, J. Maldacena, S. Minwalla, 
A. Strominger, ``S-duality and Noncommutative Gauge Theory,'' hep-th/0005048.



\bibitem{6} S. Jabbari, ``Open Strings in a B-field Background as Electric 
Dipoles,'' hep-th/9901080, Phys.Lett.B455 (1999) 129-134.




\bibitem{7} D. Bigatti, L. Susskind, 
``Magnetic fields, branes and noncommutative geometry,'' 
hep-th/9908056.




\bibitem{8} Z. Yin, ``A note on space noncommutativity,'' 
hep--th/9908056, Phys. Lett. B466 (1999) 234.


\bibitem{9}J. Gomis, T. Mehen, ``Space-Time 
Noncommutative Field Theories and Unitarity,'' hep-th/0005129.

\bibitem{12} A. Jevicki, T. Yoneya, 
``Time Uncertainty Principle and Conformal Symmetry in D-Particle 
Dynamics,'' hep-th/9805069, Nucl.Phys. B535 (1998) 335-348.




\bibitem{13} M. Li, T. Yoneya, 
``D-Particle Dynamics and The space/time Uncertainty Relation,'' 
hep-th/9611072, Phys.Rev.Lett. 78 (1997) 1219-1222.




\bibitem{14} T. Yoneya, ``String Theory and the Space--time 
Uncertainty Principle,'' hep-th/0004074 and references therein.



\bibitem{10} A. Hashimoto, I. Klebanov, ``Scattering of Strings 
from D--branes,'' hep--th/9611214, Nucl.Phys.Proc.Suppl. 55B (1997) 118.



\bibitem{11} S. Gukov, I. Klebanov, A. Polyakov, ``Dynamics of (N,1) 
Strings,'' hep-th/9711112, Phys.Lett.B423 (1998).






\end{thebibliography}
\end{document}